\documentclass[aps,prb,10pt,twocolumn]{revtex4-2}

\usepackage{graphicx}
\usepackage{dcolumn}
\usepackage{bm}
\usepackage[version=4]{mhchem}
\usepackage{float}

\begin{document}

\title{A single crystal study of Kagome metals \ce{U2Mn3Ge} and \ce{U2Fe3Ge}}

\author{Wanyue Lin$^{1,2}$, Yuchen Wu$^{1,3}$, Christopher Broyles$^{1}$, Tai Kong$^{4}$, Sheng Ran$^{1}$}
\thanks{Corresponding author: rans@wustl.edu}
\affiliation{
\\$^1$ Department of Physics, Washington University in St. Louis, St. Louis, MO 63130, USA
\\$^2$ Department of Electrical and System Engineering, Washington University in St. Louis, St. Louis, MO 63130, USA
\\$^3$ Department of Mathematics, Washington University in St. Louis, St. Louis, MO 63130, USA
\\$^4$ Department of Physics, University of Arizona, Tucson, AZ 85721, USA
}

\date{\today}

\begin{abstract}
Single crystals of U$_2$Mn$_3$Ge and and U$_2$Fe$_3$Ge with a Kagome lattice structure were synthesized using a high-temperature self-flux crystal growth method. The physical properties of these crystals were characterized through measurements of resistivity, magnetism, and specific heat. U$_2$Fe$_3$Ge exhibits ferromagnetic ground state and Anomalous Hall Effect, and U$_2$Mn$_3$Ge demonstrates a complex magnetic structure. Both compounds exhibit large Sommerfeld coefficient, indicating coexistence of heavy Fermion behaviour with magnetism. Our results suggest that this U$_2$TM$_3$Ge (TM = Mn, Fe, Co) family is a promising platform to investigate the interplay of magnetism, Kondo physics and the Kagome lattice.

\end{abstract}

\maketitle

\section{Introduction}

Kagome lattice materials have recently emerged as a focal point in condensed matter physics due to their potential to host phenomena analogous to those observed in graphene~\cite{ye2018massive}. Due to its unique geometry, Kagome lattice can support topologically protected flat bands and Dirac fermions as a result of destructive quantum phase interference in electron hopping paths~\cite{yin2022topological,jovanovic2022simple}. This characteristic enables the exploration of exotic properties in Kagome magnets and superconductors~\cite{ghimire2020topology}. Furthermore, the lattice's distinct geometry, featuring spins in triangular arrangements, is crucial in the study of quantum spin-liquid phases~\cite{yazyev2019upside,yan2011spin}. In these triangular configurations, the antiferromagnetic interactions prevent the spins from aligning alternately, resulting in a ground state deeply impacted by spin–orbit coupling, disorder, and interactions with adjacent atoms~\cite{balents2010spin}. This results in a complex interplay of competing states. Additionally, the Kagome lattice serves as an ideal platform for investigating the quantum Anomalous Hall Effect with intrinsic Berry curvature contribution~\cite{nagaosa2010anomalous}.

In this study we focus on a Kagome family that contains uranium, U$_2$TM$_3$Ge (TM = Fe, Mn or Co). The transition metal occupies the Kagome lattice, while uranium atom is on a triangle lattice~\cite{casanova1990thermoelectric}. Previous studies have shown that the magnetic properties of this family largely depend on the transition metal, although the magnetic moment is primarily associated with the uranium 5$f$ electrons. Among all the three compounds, U$_2$Fe$_3$Ge has been studied most extensively. It has a ferromagnetic ground state with a Curie temperature of 55~K. Single crystals have been synthesized using a modified Czochralski method, and magnetic properties have been measured along different crystal axes~\cite{dhar2008structure, henriques2008evidence}. The $^{57}$Fe Mössbauer spectrum demonstrates that the magnetic moments are solely from uranium~\cite{henriques2013unusual}. Due to the short distance of U-U atoms (2.799 Å and 3.2 Å), this is an exception to Hill’s rule~\cite{Hill1970}. On the other hand, U$_2$Mn$_3$Ge and U$_2$Co$_3$Ge have not been synthesized in single crystalline forms. Measurements on polycrystalline samples haven shown a Pauli paramagnetic behaviour without magnetic order for U$_2$Co$_3$Ge ~\cite{soude2011characterization}. No magnetic measurement has been reported for U$_2$Mn$_3$Ge, even on polycrystalline samples~\cite{hoffmann2013ternary}.
 
We synthesized single crystals of U$_2$Fe$_3$Ge and U$_2$Mn$_3$Ge using self flux method and characterized their magnetic properties using a combination of magnetization, electric transport and specific heat measurements. U$_2$Fe$_3$Ge has a ferromagnetic ground state with pronounced anisotropy, whereas U$_2$Mn$_3$Ge seems to have a complicated magnetic structure, likely due to the geometry frustration of the Kagome and triangle lattice. Both compounds exhibit large Sommerfeld coefficient, indicating heavy Fermion behaviour. Our results suggest that this U$_2$TM$_3$Ge (TM = Mn, Fe, Co) family is promising platform to investigate the interplay of magnetism, Kondo physics and the Kagome lattice.

\section{Experimental Methods}

Due to the incongruent melting nature and the relatively high melting temperatures, it is not easy to make high quality single crystals for a lot of uranium rich compounds, which hinders the detailed investigation of their physical properties. We used a high melting flux method, taking advantage of the deep eutectic point in
the uranium-transition metal binary phase diagrams, to synthesize single crystals of U$_2$Fe$_3$Ge and U$_2$Mn$_3$Ge.

To synthesize single crystals of U$_2$Fe$_3$Ge and U$_2$Mn$_3$Ge, specific starting stoichiometries are used. For U$_2$Fe$_3$Ge, the stoichiometry is U:Fe:Ge = 60:33:7, and for U$_2$Mn$_3$Ge, it is U:Mn:Ge = 61:33:7. The elements were first arc melted into a precursor, which was then placed into a crucible. The crucible, along with the catching crucible and a bundle of quartz wool in it acting as a filter during flux removal, was sealed into a vacuum quartz ampule. The entire ampule was heated to 1180°C in 6 hours and held for 6 hours, followed by a 100-hour cooling period to 770°C to promote crystal growth. After crystal growth, the ampule was spun in a centrifuge, which removed the still-liquid flux through the quartz wool, leaving the crystals in the crucible.

\begin{figure} 
    \centering     
    \includegraphics[width=0.45\textwidth]{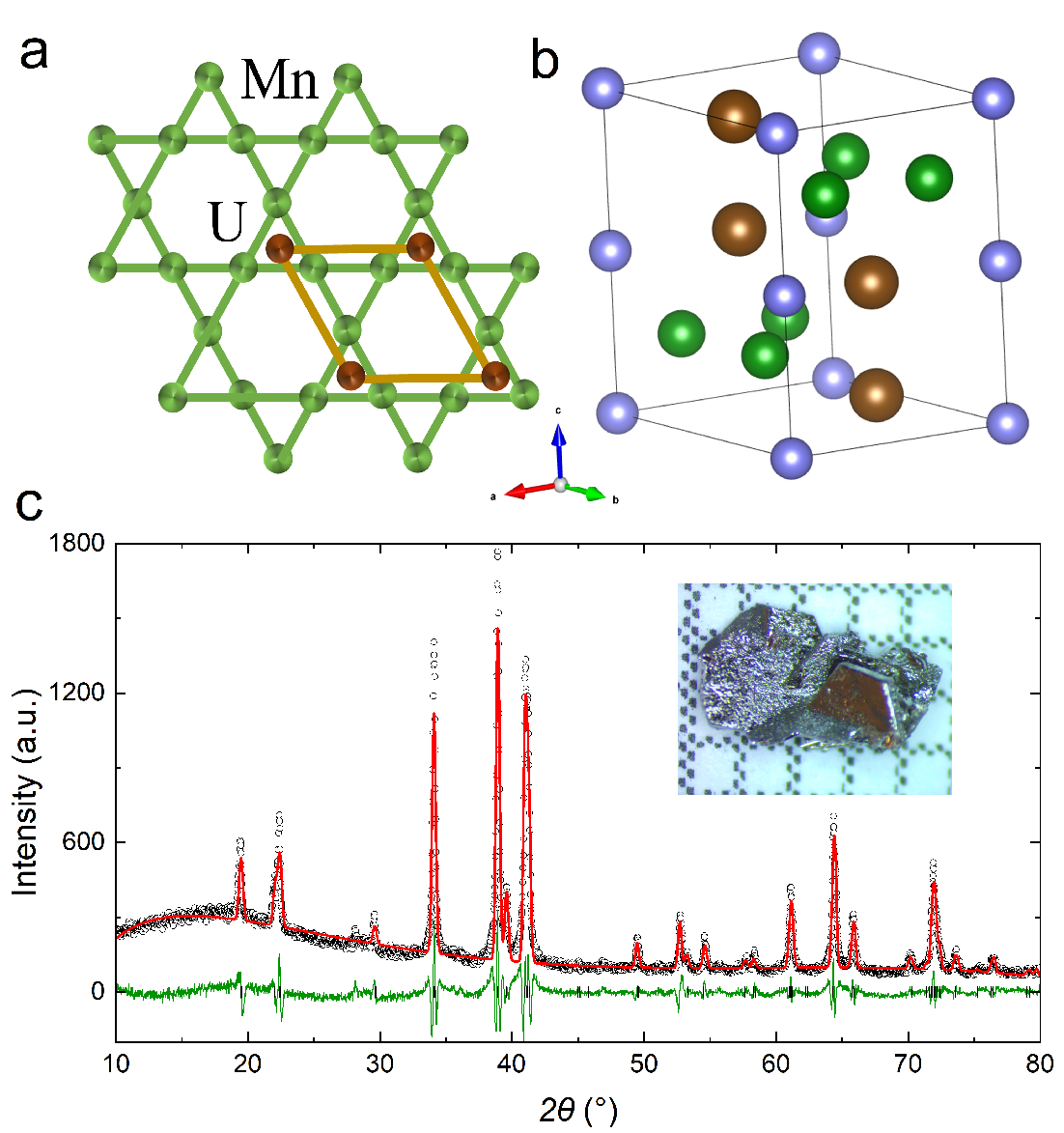} 
    \caption{(a) Schematic drawing of the Kagome and triangular plane of U$_2$Mn$_3$Ge. (b) crystal structure of U$_2$Mn$_3$Ge. Uranium, Manganese, and Germanium are represented by brown, green, and purple colors, respectively.  
    (c) The powder X-ray diffraction profile of U$_2$Mn$_3$Ge. The inset displays the as-grown single crystals, positioned on a millimeter grid paper for scale reference. }
    \label{structure} 
\end{figure}

The crystal structure was determined by $X$-ray powder diffraction using a Rigaku MiniFlex 600 diffractometer with Cu-K$_{\alpha}$ radiation ($\lambda$ = 1.5406 Å). Electrical transport, magnetization and specific heat measurements were performed in a Quantum Design Physical Property Measurement System (PPMS).

\section{Results and Discussions}
\subsection{U$_2$Fe$_3$Ge}

\begin{figure}
    \centering
    \includegraphics[width=0.45\textwidth]{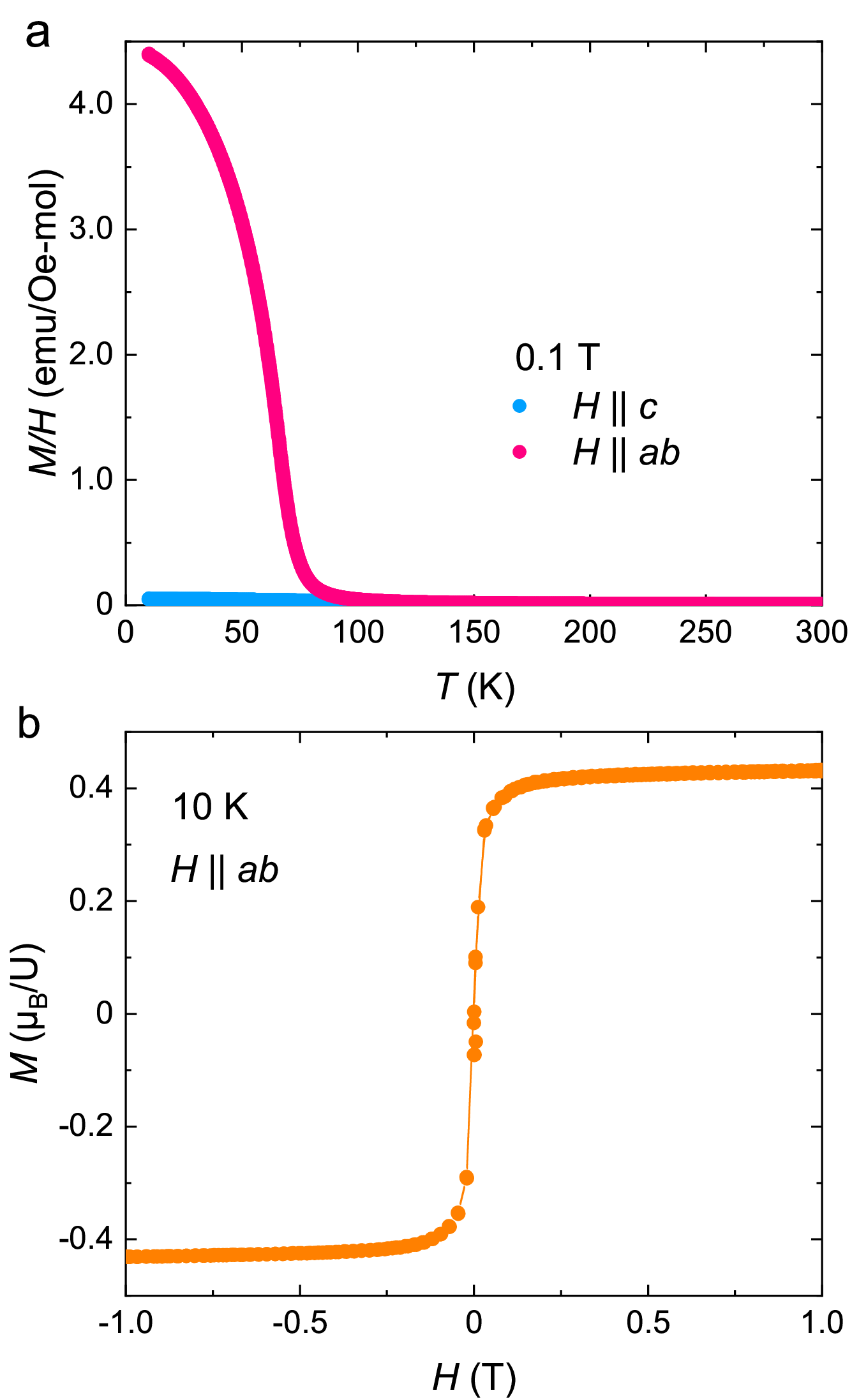}
    \caption{(a) Magnetization of U$_2$Fe$_3$Ge as a function of temperature in an external magnetic field of 0.1~T. (b) Magnetization of U$_2$Fe$_3$Ge as a function of magnetic field at 10~K. Magnetic field is applied within the $ab$ plane.}
    \label{FeMT}
\end{figure}

Previous measurements have shown that U$_2$Fe$_3$Ge has a ferromagnetic ground state with a Curie temperature of 55~K and a spontaneous magnetic moment of 0.5~$\mu_B$/U~\cite{dhar2008structure, henriques2008evidence}. The effective moment from the high temperature Curie-Weiss fit is about 3.55~$\mu_B$. This moment is larger than the free moment of U$^{3+}$ and U$^{4+}$, which seems to indicate that either Fe atom or U orbital contributes to the magnetic moment. To compare, we measured the magnetization of our single crystals. A magnetic field of 1000~Oe is applied either along the $c$ axis or within the $ab$ plane, as shown in Fig.~\ref{FeMT}. At high temperatures, the magnetization in both directions follows the the Curie-Weiss behavior. The Curie-Weiss fit to the polycrystaline average yields a magnetic moment of 2.91 ~$\mu_B$, slightly smaller than that of U$^{3+}$ and U$^{4+}$. The Curie-Weiss temperature is 75~K and 46~K in $ab$ and $c$ directions respectively. Magnetization significantly deviates from the Curie-Weiss behavior below 70~K, showing a pronounced anisotropy. The sharp increase in magnetization signifies the emergence of a ferromagnetic phase at low temperatures. The Curie temperature $T_c$, about 70~K, is higher than the previous report ~\cite{dhar2008structure}. Below $T_c$, the spontaneous magnetic moment mainly lies within the $ab$ plane. We also measured the magnetization as function of magnetic field within the $ab$ plane. The spontaneous magnetic moment is about 0.45~$\mu_B$/U. There is a very small coercive field, indicating that it is a soft ferromagnet. Note that our $M/H$ data shows very large anisotropy, with $M_{ab}/M_c$ of approximately 88. Unlike previous studies which report a magnetization ratio of approximately 1.5\cite{henriques2013unusual}, indicating mixed directional properties, our results exhibit distinct characteristics in each direction, signifying a good quality of single crystals.

\begin{figure}
    \centering
    \includegraphics[width=0.45\textwidth]{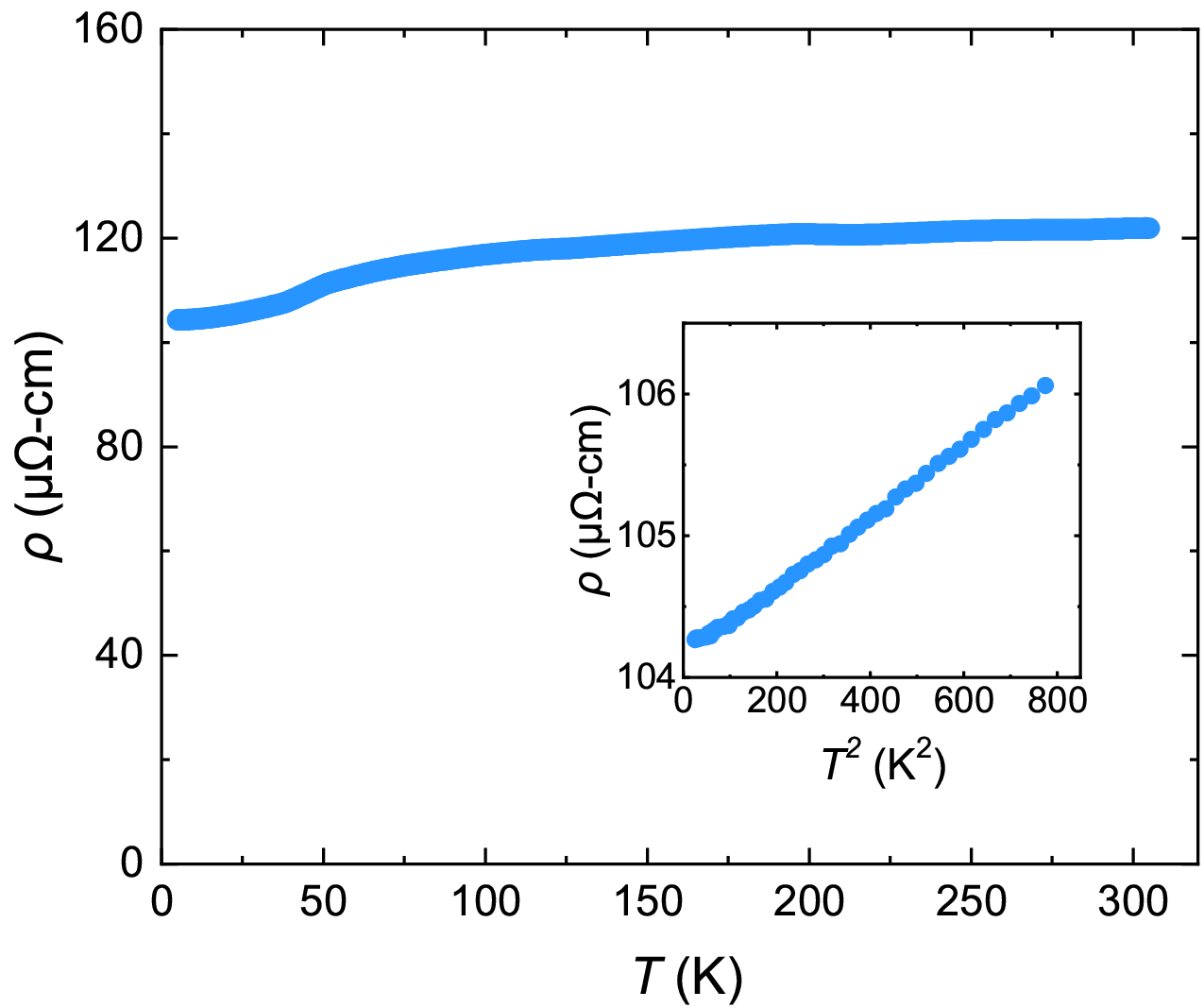}
    \caption{The electrical resistivity of U$_2$Fe$_3$Ge as a function of temperature. }
    \label{FeR}
\end{figure}

Fig. \ref{FeR} presents the temperature dependent resistivity of U$_2$Fe$_3$Ge. Consistent with prior studies, the resistivity displays a weak temperature dependence with a slight concave curvature. This behavior is typical in enhanced paramagnets and weakly ferromagnetic materials exhibiting spin fluctuations~\cite{dhar2008structure}. At low temperatures, a Fermi liquid behaviour, with $\rho \sim T^2$, is recovered. While the ferromagnetic phase has bulk nature, as confirmed by bulk magnetization measurements and previous Mössbauer measurement~\cite{henriques2013unusual}, the resistivity does not show any distinct signature for phase transition. Similar behavior has been observed in other uranium compounds, such as UFe$_2$~\cite{hvrebik1980onset}. 

\begin{figure*}
    \centering
    \includegraphics[width=0.8\textwidth]{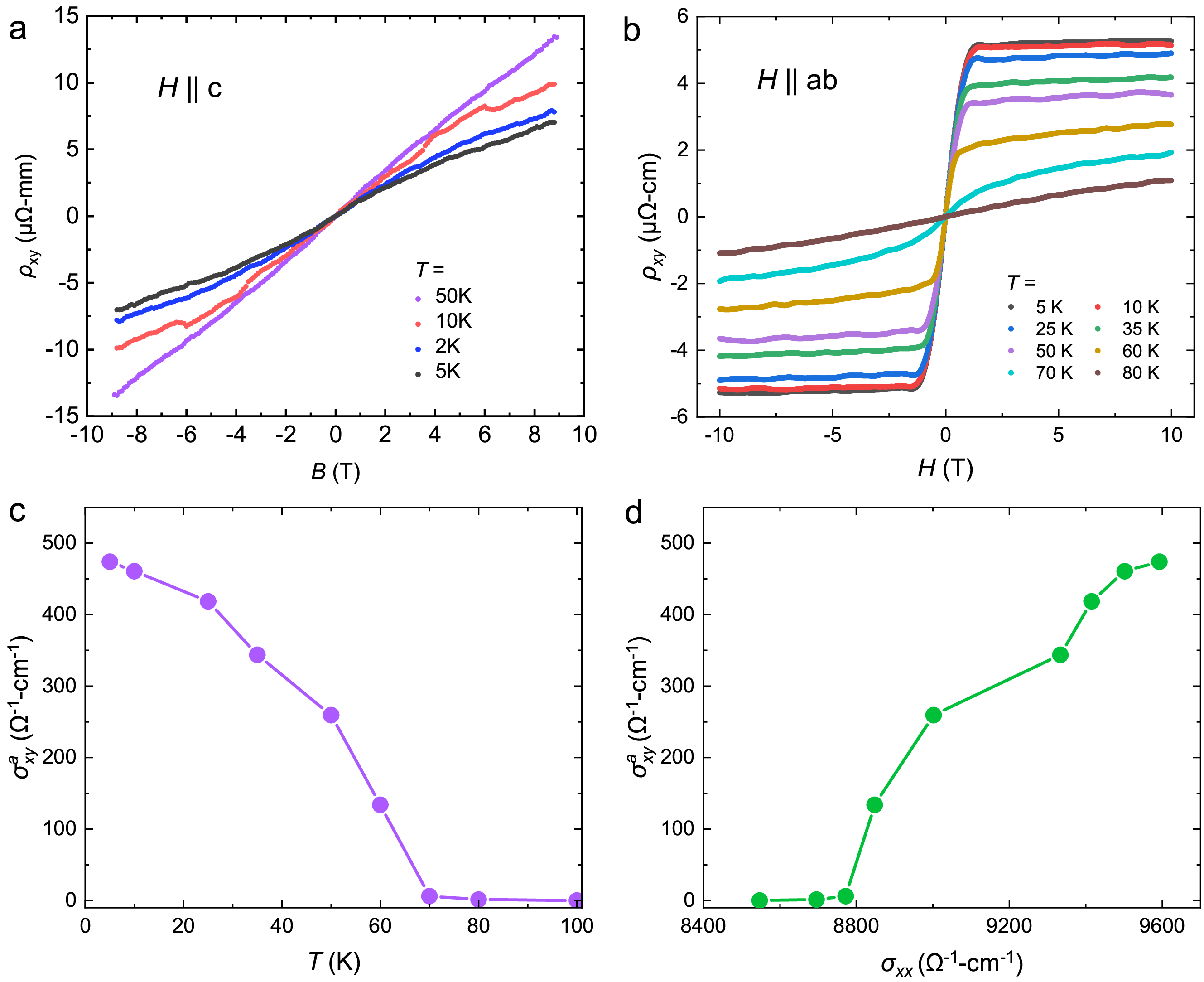}
    \caption{(a) Hall resistivity of U$_2$Fe$_3$Ge as function of magnetic field along the $c$ axis (b) Hall resistivity of U$_2$Fe$_3$Ge as function of magnetic field along the $ab$ plane (c) Anomalous Hall conductivity of U$_2$Fe$_3$Ge as a function of temperature (d) Anomalous Hall conductivity of U$_2$Fe$_3$Ge as a function of conductivity, estimated contributions to AHE from skew scattering.}
    \label{fig:hr}
\end{figure*}

Fig.\ref{fig:hr}a and \ref{fig:hr}b depict the Hall resistivity of U$_2$Fe$_3$Ge as a function of the magnetic field along the $c$ axis and $ab$ plane at various temperatures. In both directions, the Hall resistivity has a positive slope, indicating hole-type carriers. No anomalous Hall signature is observed along the $c$ axis, while a clear indication of anomalous Hall behavior is evident along the $ab$ plane, consistent with magnetization measurement. The anomalous Hall conductivity at base temperature reaches 550 $\Omega^{-1}$ cm$^{-1}$, comparable to values observed in ferromagnetic Weyl semimetals~\cite{liu2018giant}. 

The Anomalous Hall Effect (AHE) can manifest through two distinct mechanisms: extrinsic processes triggered by scattering effects and an intrinsic mechanism linked to the Berry curvature associated with the Bloch waves of electrons \cite{nagaosa2010anomalous,fang2003anomalous}. Scaling analysis has been developed and widely used in various systems to distinguish different mechanisms~\cite{liu2018giant,kim2018large,li2020giant,Siddiquee2023}. Skew scattering depends on the scattering rate, leading to a quadratic dependence of anomalous Hall conductivity on the longitudinal conductivity, $\sigma_{xy}$ $\sim$ $\sigma_{xx}^2$, while $\sigma_{xy}$ originating from the intrinsic mechanism is usually scattering-independent, and therefore, independent of $\sigma_{xx}$ and temperature. As shown in Fig.\ref{fig:hr} c and d, $\sigma_{xy}^a$ significantly depends temperature and longitudinal conductivity $\sigma_{xy}$, indicating the AHE is likely related to scattering effect. 

\subsection{U$_2$Mn$_3$Ge}

\begin{figure}
  \centering
  \includegraphics[width=0.45\textwidth]{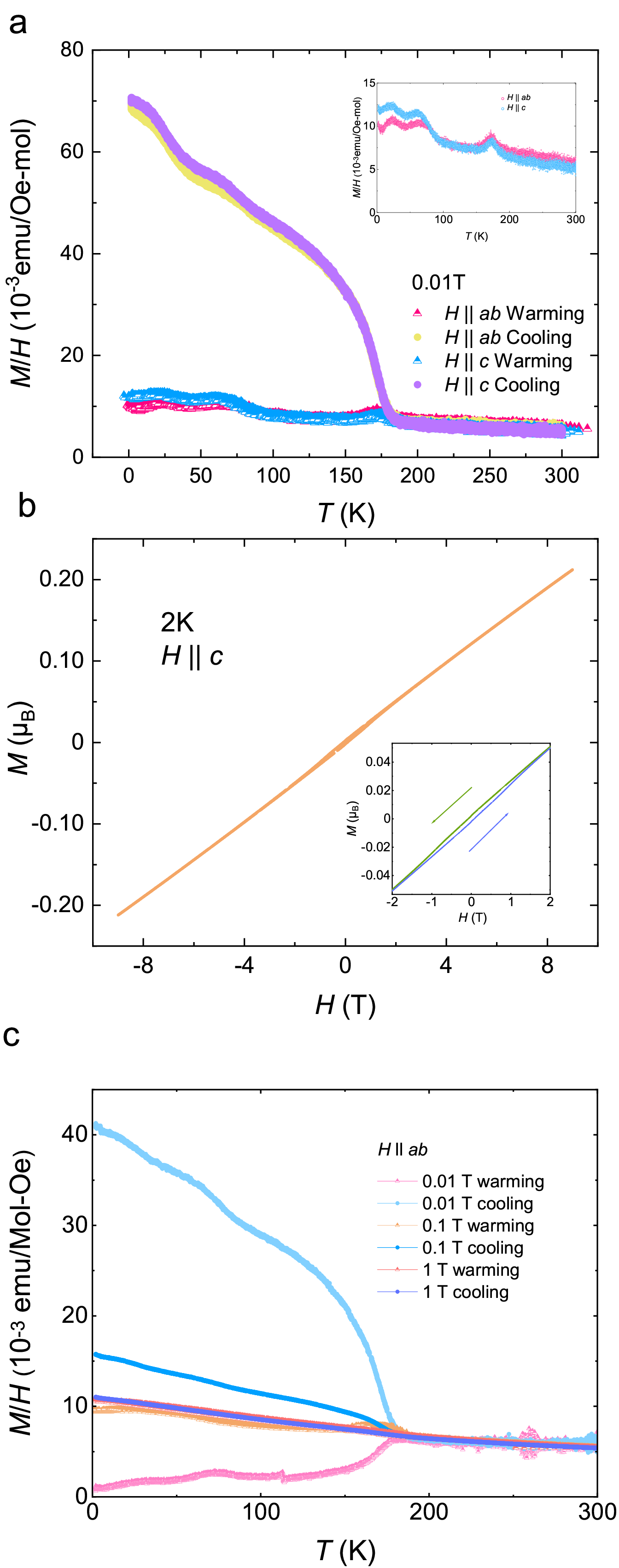}
  \caption{(a) Magnetization as a function of temperature in an external magnetic field of 0.1~T for U$_2$Mn$_3$Ge. The inset is a closer look for measurements upon warming. (b) Magnetization of U$_2$Mn$_3$Ge as a function of magnetic field at 2~K. Magnetic field is applied along the $c$ plane. The inset is a closer look for the range -2~T to 2~T. (c) Magnetization of U$_2$Mn$_3$Ge as a function of temperature under external magnetic fields of 0.01~T, 0.1~T, and 1~T.}
  \label{MnM}
\end{figure}

Polycrystalline samples of U$_2$Mn$_3$Ge have been previously synthesized~\cite{hoffmann2013ternary}, but no magnetic measurement has been reported. We measured the magnetization of the single crystals for the magnetic field along both $c$ axis and $ab$ plane, as shown in Fig.~\ref{MnM}. In sharp contrast with U$_2$Fe$_3$Ge, the magnetization shows very small anisotropy for the whole temperature range. At high temperatures, the magnetization follows the the Curie-Weiss behavior. The Curie-Weiss fit yields a magnetic moment of 2.7~$\mu_B$, which is smaller than that of U$^{3+}$ and U$^{4+}$, but comparable to that of many uranium compounds. The Curie-Weiss temperature in both directions are negative, -21~K for $c$ direction and -34~K for the $ab$ plane, indicating antiferromagnetic interaction.

As temperature decreases, a phase transition with notable hysteresis becomes apparent in the magnetization data. Upon warming, a distinct cusp in magnetization occurs at approximately 170~K, indicative of antiferromagnetic phase at low temperatures. Conversely, upon cooling, the magnetization sharply increases below 170~K, similar to a transition to ferromagnetic state. In addition, there are two broad anomalies in magnetization at lower temperatures, both upon warming and cooling, indicating the presence of multiple phase transitions. The magnetic field dependence of magnetization data at base temperature is presented in Fig.~\ref{MnM} b, with magnetic field along the $c$ axis. The magnetic moment more or less linearly increases with magnetic field. There is a small hysteresis, indicating a small amount of ferromagnetic component. Given the small ordered moment, with the upper limited of 0.02 $\mu_B$, the ferromagnetism could come from impurities. 

To gain further insights into the nature of the low temperature magnetic phases, we measured magnetization in different applied magnetic fields, as shown in Fig.~\ref{MnM}c. The hysteresis in magnetization is greatly suppressed even in magnetic field of 0.1~T, and totally disappears in 1~T. It is possible that there exists competing magnetic interactions. This is consistent with the fact that the Curie-Weiss temperature is significantly smaller than the magnetic phase transition temperature. In this case, a small magnetic field can easily stabilize a different magnetic phase. Alternatively, the magnetic phase transition could come from small amount of impurity phase, which is aligned in a small magnetic field. In this case, U$_2$Mn$_3$Ge itself does not have any magnetic order, despite of the large finite free magnetic moment and Curie-Weiss temperature. The absence of the magnetic order could be due to the geometry frustration of the Kagome and triangle lattice. In any of these two cases, U$_2$Mn$_3$Ge provides a good platform to investigate the frustrated magnetism and competing magnetic interactions. 

\begin{figure}
  \centering
   \includegraphics[width=0.45\textwidth]{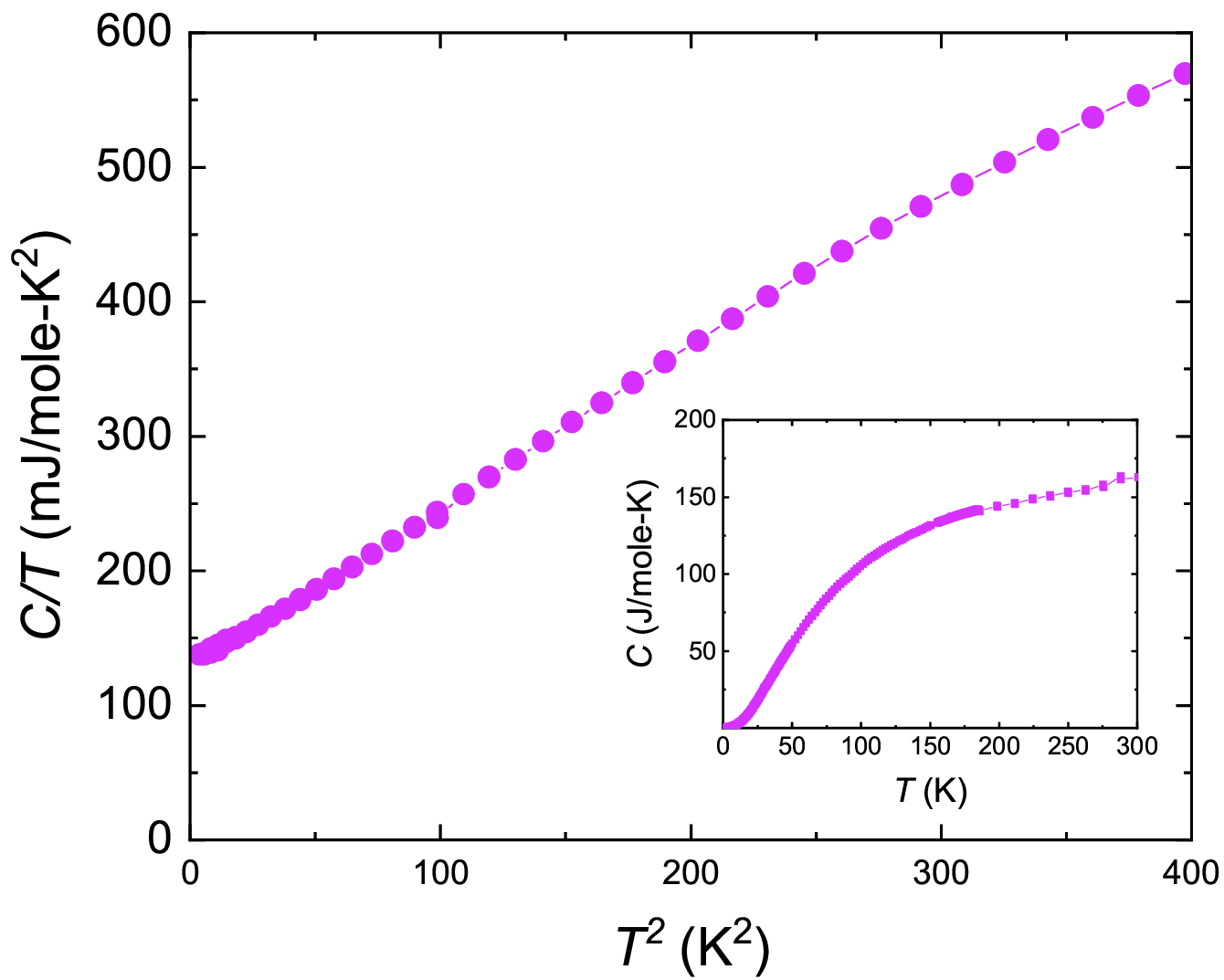}
  \caption{Specific heat divided by temperature \(C/T\) as a function of \(T^2\) at low temperatures, showing the electronic contribution to the specific heat. The inset shows specific heat \(C\) as a function of temperature \(T\) in the full temperature range.}
  \label{Mnsp}
\end{figure}

\begin{figure}
    \centering 
    \includegraphics[width=0.45\textwidth]{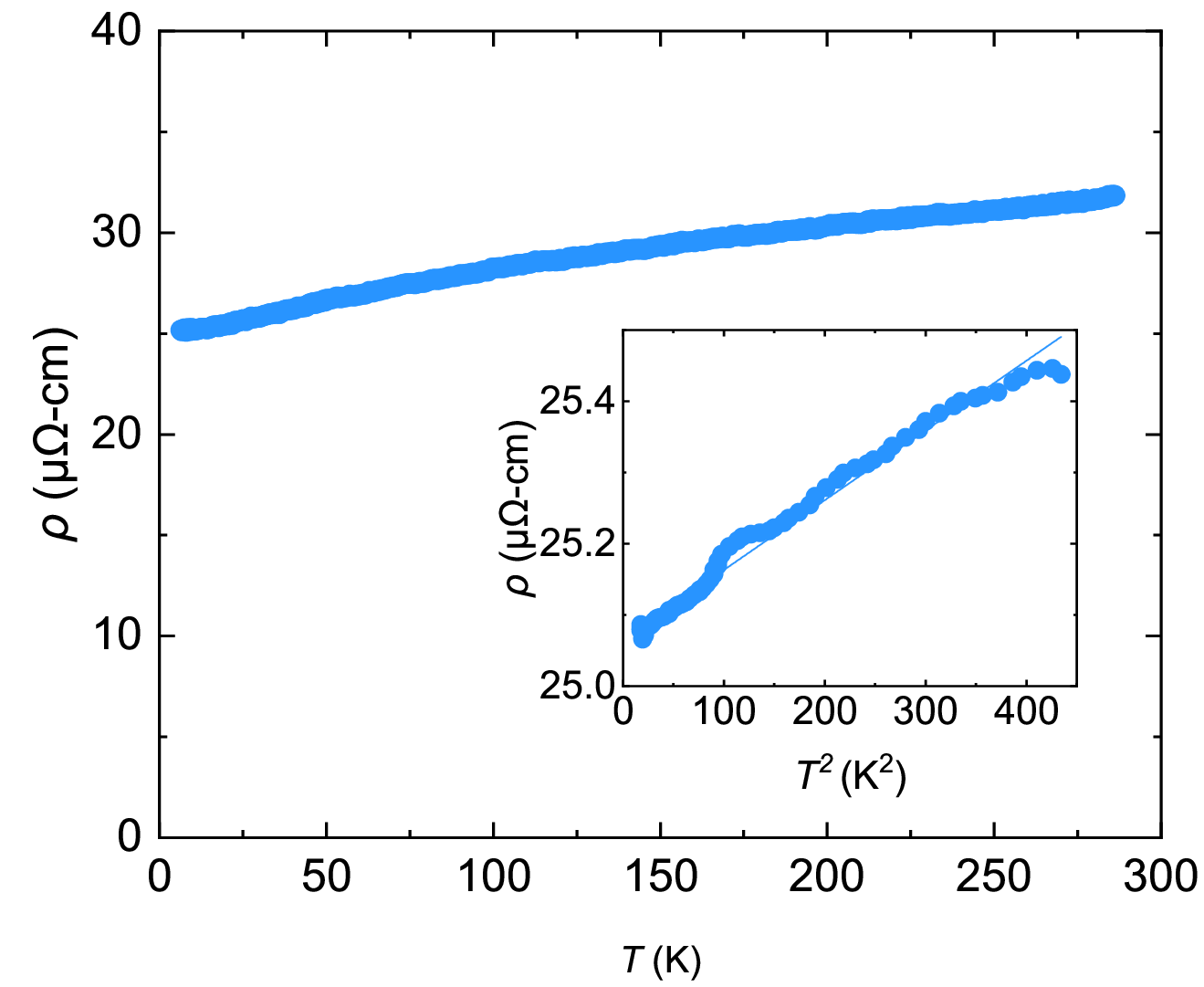}
    \caption{The electrical resistivity of U$_2$Mn$_3$Ge as a function of temperature.} 
    \label{Mn Resistivity} 
\end{figure}

Similar to U$_2$Fe$_3$Ge, the phase transition does not exhibit in either resistivity or specific heat data shown in  Fig. \ref{Mn Resistivity} and Fig. \ref{Mnsp}. Resistivity has a concave curvature like U$_2$Fe$_3$Ge, but less pronounced. Fermi liquid behaviour is also recovered at low temperatures. We want to emphasize that the absence of signature of phase transition in resistivity and specific heat does not indicate the absence of bulk magnetic order. As mentioned above, in U$_2$Fe$_3$Ge, the bulk magnetic order is evidenced in Mössbauer measurement~\cite{dhar2008structure}, and yet no noticeable anomalies are observed in resistivity and specific heat~\cite{dhar2008structure,henriques2013unusual}. While impurities in the sample could be a potential cause, our X-ray diffraction pattern does not show obvious peaks from impurities. A plausible explanation for this could be the inherent complexities of frustrated systems, which might result in only a small change in entropy during the transition. In this case, the dominant phonon heat capacity could overshadow the subtle magnetic effects, making these transitions too subtle to be detected. 

To extract the electronic contribution to the specific heat, we approximated $C/T$ with $\gamma +\beta T^2$. Our observed value for \( \gamma \) is around 128~mJ/mol-K$^2$, which is considerably large. Previous heat capacity measurements of U$_2$Fe$_3$Ge also yield a large Sommerfeld coefficient \( \gamma \) of $88 - 97.8 \, \text{mJ mol}^{-1} \text{K}^{-2}$~\cite{dhar2008structure,henriques2013unusual}. Such high values of the Sommerfeld coefficient suggest a coexistence of Kondo lattice and magnetic ordered state. In a standard Kondo lattice system, the local Kondo interactions are in competition with Ruderman-Kittel-Kasuya-Yosida (RKKY) interactions, which tends to stabilize long-range magnetic order. However, uranium compounds often possess multiple 5$f$ electrons, in stark contrast to cerium-based 4$f$ Kondo lattice systems, which usually have a single 4$f$ electron. Existence of multiple electrons in uranium compounds introduces a novel aspect: the potential for an orbital-selective coexistence of magnetism and Kondo lattice behavior~\cite{Giannakis2019,Chen2019,Siddiquee2023}. Here, specific 5$f$ electrons may contribute to magnetic ordering, while others engage in Kondo interactions. Given that both U$_2$Mn$_3$Ge and U$_2$Fe$_3$Ge show a large Sommerfeld coefficient, it is likely a common feature in this family of Kagome lattice systems, which makes it a  promising platform to investigate the coexistence of Kondo physics, magetism and the Kagome lattice.

\section{Conclusion}
In conclusion, we have synthesized single crystals of Kagome lattice compounds U$_2$Mn$_3$Ge and U$_2$Fe$_3$Ge, and studied their magnetic, electric transport and thermal properties. U$_2$Fe$_3$Ge has a ferromagnetic ground state with pronounced anisotropy, whereas U$_2$Mn$_3$Ge seems to have a complicated magnetic structure, likely due to the geometric frustration of the Kagome and triangle lattice. Both compounds exhibit a large Sommerfeld coefficient, indicating heavy Fermion behaviour. Moreover, a clear indication of anomalous Hall behavior is evident for magnetic field along the $ab$ plane in U$_2$Fe$_3$Ge. Our results suggest that this U$_2$TM$_3$Ge (TM = Mn, Fe, Co) family is a promising platform to investigate the interplay of magnetism, Kondo physics and the Kagome lattice. 

\section{Acknowledgments}

Research at Washington University is supported by the National Science Foundation (NSF) Division of Materials Research Award DMR-2236528. Research at the University of Arizona is based upon work supported by the National Science Foundation under Award No. DMR-2338229.

\section{Data availability statement}

Any data that support the findings of this study are included within the article.

\nocite{*}

\bibliography{reference}
\end{document}